 \documentclass[prl,twocolumn,floatfix,superscriptaddress]{revtex4}
\usepackage{dcolumn,amsmath}
\usepackage{bm}
\usepackage{hyperref}   

\def\veps{\varepsilon}

\newcommand{\tref}[1]{Table~\ref{#1}}

\begin{document}
\title{Towards the electron EDM search: Theoretical study of HfF$^+$}
\author{A.N.\ Petrov}\email{anpetrov@pnpi.spb.ru}
\altaffiliation [Also at ] {St.-Petersburg State University, St.-Petersburg,
        Russia}
\author{N.S.\ Mosyagin}
\author{T.A.\ Isaev}
\author{A.V.\ Titov} \altaffiliation [Also at ]
{St.-Petersburg State University, St.-Petersburg, Russia}
\affiliation
{Petersburg Nuclear Physics Institute, Gatchina,
             Leningrad district 188300, Russia}


\begin{abstract}
 We report first {\it ab initio} relativistic correlation calculations of
 potential curves for ten low-lying electronic states, effective electric field
 on the electron and hyperfine constants for the $^3\Delta_1$ state of cation
 of a heavy transition metal fluoride, HfF$^+$, that is suggested to be used
 as the working state in experiments to search for the electric dipole moment
 of the electron. It is shown that HfF$^+$ has deeply bound $^1\Sigma^+$
 ground state, its dissociation energy is $D_e=6.4$ eV.
 The $^3\Delta_1$ state is obtained to be the relatively long-lived first
 excited state lying about 0.2 eV 
 higher. The calculated effective electric field $E_{\rm eff}=W_d|\Omega|$
 acting on an electron in this state is $5.84\times10^{24}{\rm
 Hz}/(e{\cdot}{\rm cm})$.
\end{abstract}

\maketitle

 \paragraph*{Introduction.}
%
 The search for the electric dipole moment (EDM) of the electron, $d_e$ or eEDM below,
 remains one of the most fundamental problems in physics.
 Up to now only upper limits were obtained for $|d_e|$. The tightest bound on
 $d_e$ was obtained in the experiment on the Tl atom \cite{Regan:02}, which
 established an upper bound of $|d_e|<1.6\times10^{-27}\ e\cdot$cm ($e$ is the
 charge of the electron). It seems unlikely that experiments on atoms can give
 tighter bounds.
 Molecular systems, on the other hand, provide a way for much
 enhanced sensitivity, since the effective intramolecular electric field acting
 on electrons in polar molecules can be five or more orders of magnitude higher
  than the maximal field available in the laboratory \cite{Sushkov:78,
  Gorshkov:79, Titov:06amin}.

 The new generation of the eEDM experiments, employing polar heavy-atom
 molecules, is expected to reach sensitivity of $10^{-30}{-}10^{-28}
 e{\cdot}{\rm cm}/\sqrt{\textrm{day}}$ (e.g., see \cite{Sauer:06a}).  Their
 results are expected to dramatically influence all the popular extensions of
 the standard model, in particular supersymmetry, even if bounds on the
 $P,T$-odd effects compatible with zero are obtained (see \cite{Ginges:04,
 Erler:05} and references therein). These studies include the beam experiments
 carried out on YbF molecular radicals by E.~Hinds group at Imperial College,
 London \cite{Sauer:06a}, and the vapor cell experiment on the metastable
 $a(1)$ state of PbO prepared by the group of DeMille at Yale University (see
 \cite{Kawall:04b} and references therein). New ways of searching for eEDM,
 using trapped cold molecular cations, were investigated during last two years
 at JILA by E.~Cornell group.  The first candidate was HI$^+$ \cite{Stutz:04},
 but subsequent estimate \cite{Ravaine:05a} and accurate calculations
 \cite{Isaev:05a} have showed that the cation has rather small $E_{\rm eff}$
 (see below).
 Then some more motivated candidates were considered, HfH$^+$
 \cite{Sinclair:05Aa}, HfF$^+$ \cite{Cornell:06PC} etc., having, possibly,
 working $^3\Delta_1$ as the ground or at least long-lived first excited state.
 Even experimental study of spectroscopic properties for this cation is a very
 difficult problem and there are no such data measured up to date though they
 are required already to analyze basic stages of the eEDM experiment
 \cite{Cornell:06PC}. In turn, the modern relativistic computational methods
 can now give reliable answers on almost all the questions of interest even for
 compounds of heavy transition metals as is in our case.  So, the goal of the
 present paper is a theoretical study of required spectroscopic properties,
 $E_{\rm eff}$ and hyperfine structure of HfF$^+$ as a candidate for the eEDM
 experiment. Such a complete analysis is the first one performed for the given
 kind of molecules.

\paragraph*{$E_{\rm eff}$, hyperfine structure and working state in HfF$^+$.}
%
 One of the most important features of such experiments is that the knowledge
 of the effective electric field, $E_{\rm eff}$, seen by an unpaired electron
 is required for extracting $d_e$ from the measurements. $E_{\rm eff}$ can not
 be obtained in an experiment, rather, electronic structure calculations are
 required for its evaluation.  It is presented by the expression
 $E_{\rm eff}=W_d|\Omega|$, where $W_d$ is a parameter of the P,T-odd molecular
 Hamiltonian that is given in Refs.\,\cite{Kozlov:87, Kozlov:95, Titov:06amin}

\begin{equation}
   W_d = \frac{1}{\Omega d_e}
   \langle \Psi|\sum_iH_d(i)|\Psi \rangle~,
\end{equation}
 where $\Psi$ is the wavefunction for the working $^3\Delta_1$ state, and
 $\Omega= \langle\Psi|\bm{J}\cdot\bm{n}|\Psi\rangle = \pm1 $, $\bm{J}$ is the
 total electronic momentum, $\bm{n}$ is the unit vector along the molecular
 axis directed from Hf to F,

\begin{eqnarray}
    H_d=2d_e
    \left(\begin{array}{cc}
    0 & 0 \\
    0 & \bm{\sigma E} \\
    \end{array}\right)\ ,
\label{Wd}
\end{eqnarray}
 $\bm{E}$ is the inner molecular electric field, $\bm{\sigma}$ are the Pauli
 matrices. Recently, the $E_{\rm eff}$ value for the $^3\Delta_1$ state was
 estimated in the scalar-relativistic approximation by Meyer {\it et al.}
 \cite{Meyer:06a}
 and the method used for calculation of $E_{\rm eff}$ is close in essence to
 that developed by us earlier \cite{Titov:85Dis} and applied to first two-step
 calculations of the PbF molecule \cite{Titov:85Dis, Kozlov:87}.  In the
 present work, a more reliable value of $E_{\rm eff}$ is calculated using the
 advanced two-step techniques developed by us later \cite{Titov:96,
 Mosyagin:98, Petrov:02, Petrov:05a}.
%
 The hyperfine constants for $^{177}$Hf and $^{19}$F nuclei
 ($A_{\parallel}[\rm{Hf}]$ and $A_{\parallel}[\rm{F}]$) for the $^3\Delta_1$
 state of HfF$^+$ are also calculated.  When the experimental value
 $A_{\parallel}[\rm{Hf}]$
   is measured, it will provide
 an accuracy check for the calculated $E_{\rm eff}$ value. Both
 $A_{\parallel}[\rm{Hf}]$ and $A_{\parallel}[\rm{F}]$ values are useful for
 identifying HfF$^+$ by his spectrum from another species in the experiment.
 Moreover, preparation of the HfF$^+$ cations in the working state for the eEDM
 experiment, registration of eEDM signals etc.\ requires the knowledge of
 spectroscopic properties. In particular, an open question is to clarify, which
 state is the ground one. If $^3\Delta_1$ is not the ground state, its
 radiative lifetime is also required.  In the present work these data are
 obtained, though precise studying of the compounds with transition elements is
 a difficult problem for modern molecular theory.  The needed spectroscopic
 information cannot be obtained now from other sources.

 \paragraph*{Methods and calculations.}
%
 The 12-electron generalized relativistic effective core potential (GRECP)
 \cite{Titov:00a} simulating interaction with the explicitly excluded $1s$ to
 $4f$ electrons of Hf is used in 20-electron calculations of HfF$^+$. In
 10-electron calculations which are substantially less time consuming, $5s$ and
 $5p$ spinors of hafnium and $1s$ orbital of fluorine are frozen 
  from the states averaged over the nonrelativistic configurations
  $5d^2 6s^{0.6} 6p^{0.4}$ for Hf$^+$ and $2s^2 2p^5$ for F
 and not treated explicitly. The generalized correlation atomic basis set
 (GCBS) \cite{Mosyagin:00,Isaev:00} ($12s16p16d10f10g$)/[$6s5p5d3f1g$] is
 constructed for Hf. The ANO-L ($14s9p4d3f$)/[$4s3p2d1f$] atomic basis set
 listed in the {\sc molcas~4.1} library \cite{MOLCAS} was used for fluorine.
 The molecular orbitals are obtained by the complete active space
 self-consistent field (CASSCF) method \cite{Olsen:88,MOLCAS} with the
 spin-averaged part of the GRECP \cite{Titov:99}, i.e.\ only
 scalar-relativistic effects are taken into account at this stage.  In the
 CASSCF method, orbitals are subdivided into three groups: inactive, active and
 virtual. Inactive orbitals are doubly occupied in all the configurations, all
 possible occupations are allowed for active orbitals, whereas virtual orbitals
 are not occupied.  So, wave function is constructed as a full configuration
 interaction expansion in the space of active orbitals and both active and
 inactive orbitals are optimized for subsequent correlation calculations of
 HfF$^+$. Using the $C_{2v}$ point group classification scheme, five orbitals
 in A$_1$, four in B$_1$ and B$_2$ and two in A$_2$ irreps are included into
 the active space.  In 10-electron calculations, one orbital in the A$_1$
 irreps (which is mainly $2s$ orbital of F) belongs to the inactive space. In
 20-electron CASSCF calculations, the $5s$ and $5p$ orbitals of Hf and $1s$
 orbital of F are added to the space of inactive orbitals.

 Next, the spin-orbit direct configuration interaction (SODCI) approach
 \cite{Buenker:74, Buenker:99, Alekseyev:04a} modified by our group to account
 for spin-orbit interaction in configuration selection procedure
 \cite{Titov:01} with the selected single and double excitations from some
 multiconfigurational reference states (``mains'') is employed on the sets of
 different {$\Lambda$}S many-electron spin- and space-symmetry adapted basis
 functions (SAFs).  Details on features of constructing the reference space and
 selection procedure are given in Refs.~\cite{Titov:01, Mosyagin:02,
 Petrov:05a}.

 Ten lowest states with the leading configurations
 $[...]\sigma_1^2\sigma_2^2$ ($^1\Sigma^+$),
 $[...]\sigma_1^2\sigma_2^1\delta^1$ ($^3\Delta_{1,2,3}$;~$^1\Delta$),
 $[...]\sigma_1^2\sigma_2^1\pi^1$~($^3\Pi_{0^-,0^+,1,2}$;~$^1\Pi$)
 were calculated. Here $\sigma_1$ orbital is mainly formed by $2p_z$ orbital of
 F with admixture of $6p_z$ and $6s$ orbitals of Hf, $\sigma_2$ is mainly $6s$
 orbital of Hf with admixture of $6p_z$ orbital of Hf, $\delta$ and $\pi$ are
 mainly the $5d$ orbitals of Hf.

 To obtain spectroscopic parameters, six points between 3.1 and 4.0 a.u.\ and
 point at 100 a.u.\ of the HfF$^+$ potential curves were calculated for ten
 lowest-lying states in 10-electron calculations and for four states in
 20-electron ones. The 20-electron calculation is substantially more
 time-consuming and for the rest six
 states were calculated only for one point, 3.4 a.u., in the present study.
 Comparing the latter calculations with corresponding 10-electron ones the core
 ($5s^2, 5p^6$ shells of Hf and $1s^2$ shell of F) relaxation and correlation
 corrections to the $T_e$ values, called ``core corrections'' below and in
 \tref{spec}, were estimated.  The core properties $A_{\parallel}$ and $E_{\rm
 eff}$ were calculated only for the working $^3\Delta_1$ state at point 3.4
 a.u., which is close to the equilibrium distance (see \tref{spec}).  Before
 calculating core properties the shapes of the four-component molecular spinors
 are restored in the inner core region after the two-component GRECP
 calculation of the molecule. For this purpose the nonvariational one-center
 restoration (NOCR) method \cite{Titov:85Dis, Titov:96, Titov:96b, Titov:99,
 Petrov:02, Petrov:05a} is applied.

\paragraph*{Results and discussion.}
%
 The results of calculations for HfF$^+$ spectroscopic parameters are presented
 in \tref{spec}. The first point to note is that the cation is deeply bound.
 Second important result is that the $^1\Sigma^+$ state appears to be the
 ground one and the working state, $^3\Delta_1$, is the first excited one.
 Besides, the excitation energy from $^1\Sigma^+$ to $^3\Delta_1$ is increased
 from 866 $cm^{-1}$ to 1633 $cm^{-1}$ after including the $5s$ and $5p$ shells
 of Hf and $1s$ shell of F to the relativistic correlation calculation.
 Excitation energies from $^1\Sigma^+$ to other calculated low-lying states are
 also increased. Note also that the values obtained in 10-electron calculations
 for lowest four states when just the core  correction described above is taken
 into account are in a good agreement with the purely 20-electron calculations.
 (Note that accounting for correlation/relaxation of the $4f$-shell would be
 also desirable but it is too consuming and we expect it will result in smaller
 change of spectroscopic properties than that for the above core correction.)
%
%
%
 The SO splittings of the $^3\Delta$ and $^3\Pi$ states are mainly due to the
 SO splitting of the $5d$ shell of Hf:
\begin{equation}
   {\bf H}_{ls}^{\rm so} = {\it a}\cdot({\bf l_{\rm 5d}}{\cdot}{\bf s_{\rm 5d}})\ ,
\label{hso}
\end{equation}
 The atomic Dirac-Fock calculation of Hf$^+$ gives
 $\veps_{5/2}{-}\veps_{3/2}=3173 {\rm ~cm}^{-1} \Rightarrow {\it a}=1269 {\rm
 ~cm}^{-1}$, where $\veps_{5/2}$ and $\veps_{3/2}$ are orbital energies of
 $5d_{5/2}$ and $5d_{3/2}$ states of Hf$^+$. The SO interaction (\ref{hso})
 averaged over the $^3\Delta$ or $^3\Pi$ states is reduced to
%
%
%
\begin{equation}
   {\bf H}_{LS}^{\rm so} = {\cal A}\cdot({\bf L}{\cdot}{\bf S}),
\label{hso2}
\end{equation}
%
 where ${\cal A}= 1269/2 = 635 {\rm ~cm}^{-1}$, $\bf{L}$ and $\bf{S}$ are the
 orbital and spin momenta of HfF$^+$. The SO interaction (\ref{hso2}) leads to
 splitting between components of the $^3\Delta$ and $^3\Pi$ states on 1269 and
 635${\rm ~cm}^{-1}$, correspondingly. It is in a good agreement with the
 splitting of the $^3\Delta$ state calculated in \tref{spec} but not with the
 $^3\Pi$ one because of the SO interaction with closely lying $^1\Delta$ and
 $^1\Pi$ states.

 The calculated $A_{\parallel}$ and $E_{\rm eff}$ for the $^3\Delta_1$ state
 are presented in \tref{core}. In opposite to $A_{\parallel}[\rm{F}]$,
 $A_{\parallel}[\rm{Hf}]$ and $E_{\rm eff}$ are not seriously changed when the
 outer core electrons are included into calculation. Such behavior of
 $A_{\parallel}[\rm{F}]$ is explained by the fact that fluorine can be
 considered with good accuracy as a closed shell subsystem (thus having
 $A_{\parallel}[\rm{F}]\approx0$) in the $^3\Delta$ state, i.e.\ large
 compensation of contributions from orbitals with different projections of
 total electronic momentum for $A_{\parallel}[\rm{F}]$ takes place.
 Therefore, even small perturbation can seriously change the
 $A_{\parallel}[\rm{F}]$ value. Calculated $E_{\rm eff}$ is large and
 comparable with the corresponding value for the $a(1)$ state of the PbO
 molecule \cite{Petrov:05a}.  If one does not pay attention to the difference
 of signs our value is in 1.34 times larger than value obtained in the paper
 \cite{Meyer:06a} in scalar-relativistic calculations.  Our rather crude
 estimate of the $^3\Delta_1$ lifetime for the $^3\Delta_1 \rightarrow
 {^1}\Sigma^+$ transition is about 1/2 sec.  This value is difficult for
 accurate calculations because of small absolute values of both transition
 energies and, particularly, transition dipole moments between those states
 whereas absolute errors are similar to those for other transitions.

\begin{table}
\caption
  {Calculated spectroscopic parameters for HfF$^+$} 
\begin{tabular}{ccrrcc}
\\
\hline
\hline
 \vspace{-3 mm} \\
  State  & $R_e$ $\AA$~~ & $T_e$ $cm^{-1}$ & ~~$T_e$ with core & ~~$w_e$ $cm^{-1}$ & $D_e$ $cm^{-1}$ \\
       &               &                 & correction$^a$ &                   &                 \\
\multicolumn{6}{c}{}
 \vspace{-3 mm} \\
\hline
 \vspace{-3 mm} \\
\multicolumn{5}{c}{\bf 10-electron calculation}\\
 \vspace{-3 mm} \\
\hline
 \vspace{-3 mm} \\
     { $^1\Sigma^+$ } & 1.784    &     0   &     0  &  751  & 51107 \\
     { $^3\Delta_1$ } & 1.810    &   866   &  1599  &  718  & \\
     { $^3\Delta_2$ } & 1.809    &  1821   &  2807  &  719  & \\
     { $^3\Delta_3$ } & 1.807    &  3201   &  4324  &  721  & \\
     { $^1\Delta_2$ } & 1.814    &  9246   & 11519  &  696  & \\
     { $^3\Pi_{0^-}$} & 1.856    &  9466   & 11910  &  689  & \\
     { $^3\Pi_{0^+}$} & 1.854    &  9753   & 12196  &  699  & \\
     { $^3\Pi_1  $  } & 1.860    & 10190   & 12686  &  687  & \\
     { $^3\Pi_2  $  } & 1.856    & 11898   & 14438  &  703  & \\
     { $^1\Pi_1  $  } & 1.870    & 12642   & 14784  &  679  & \\
 \vspace{-3 mm} \\
\hline
 \vspace{-3 mm} \\
\multicolumn{6}{c}{\bf 20-electron calculation}\\
 \vspace{-3 mm} \\
\hline
 \vspace{-3 mm} \\
     { $^1\Sigma^+$ } & 1.781    &     0   &        & 790  & 51685 \\
     { $^3\Delta_1$ } & 1.806    &  1633   &        & 746  & \\
     { $^3\Delta_2$ } & 1.805    &  2828   &        & 748  & \\
     { $^3\Delta_3$ } & 1.804    &  4273   &        & 749  & \\
 \vspace{-3 mm} \\
\hline
\hline

\end{tabular}
\begin{flushleft}
 $^{\rm a}$ See paragraph ``Methods and calculations'' for details.\\
\end{flushleft}
\label{spec}
\end{table}

\begin{table}
\caption{
   Calculated parameters $A_{\parallel}[\rm{Hf}]$ and
   $A_{\parallel}[\rm{F}]$ (in MHz) and $E_{\rm eff}$ (in $10^{24}{\rm
   Hz}/(e\cdot{\rm cm})$) for the $^3\Delta_1$ state of
   $^{177}$Hf$^{19}$F$^+$ at internuclear 
   distance of 3.4 a.u.
}

\begin{tabular}{ccc}
\\
\hline
\hline
 \vspace{-2 mm} \\
~~$A_{\parallel}[\rm{Hf}]$~~  & ~~$A_{\parallel}[\rm{F}]$~~~ & ~~~$E_{\rm eff}$~~ \\
 \vspace{-2 mm} \\
\hline
 \vspace{-3 mm} \\
\multicolumn{3}{c}{\bf 10-electron calculation } \\
 \vspace{-3 mm} \\
\hline
 \vspace{-2 mm} \\
-1250 &  -33.9   &  5.89  \\
 \vspace{-2 mm} \\
\hline
 \vspace{-3 mm} \\
\multicolumn{3}{c}{\bf 20-electron calculation } \\
 \vspace{-3 mm} \\
\hline
 \vspace{-2 mm} \\
 -1239     &   -58.1       &   5.84 \\
 \vspace{-2 mm} \\
\hline
\hline
\end{tabular}
\label{core}
\end{table}


\paragraph*{Acknowledgments.}
 This work is supported by the RFBR grant 06--03--33060.  T.I.\ and A.P.\ are
 also supported by grants of Russian Science Support Foundation.  A.P.\ is
 grateful for the grant of Governor of Leningrad district.

\bibliographystyle{./bib/apsrev}


\bibliography{bib/JournAbbr,bib/Titov,bib/TitovLib,bib/Kaldor,bib/Isaev}

\end{document}